# Properties of a Simple Bilinear Stochastic Model: Estimation and Predictability

D. Sornette[1] and V.F. Pisarenko[2]

[1]ETH Zurich
Department of Management, Technology and Economics
CH-8032 Zurich, Switzerland

[3]International Institute of Earthquake Prediction Theory
and Mathematical Geophysics
Russian Ac. Sci. Warshavskoye sh., 79, kor. 2, Moscow 113556, Russia

**Abstract**
We analyze the properties of arguably the simplest bilinear stochastic multiplicative process, proposed as a model of financial returns and of other complex systems combining both nonlinearity and multiplicative noise. By construction, it has no linear predictability (zero two-point correlation) but a certain nonlinear predictability (non-zero three-point correlation). It can thus be considered as a paradigm for testing the existence of a possible nonlinear predictability in a given time series. We present a rather exhaustive study of the process, including its ability to produce fat-tailed distribution from Gaussian innovations, the unstable characteristics of the inversion of the key nonlinear parameters and of the two initial conditions necessary for the implementation of a prediction scheme and an analysis of the associated super-exponential sensitivity of the inversion of the innovations in the presence of a large impulse. Our study emphasizes the conditions under which a degree of predictability can be achieved and describe a number of different attempts, which overall illuminates the properties of the process. In conclusion, notwithstanding its remarkable simplicity, the bilinear stochastic process exhibits remarkably rich and complex behavior, which makes it a serious candidate for the modeling of financial time series among others.

## 1. Introduction

Daily asset returns in liquid markets exhibit two key statistical properties: (i) price changes are not auto-correlated beyond a few minutes, but (ii) the absolute values of changes are auto-correlated over long time scales leading to long-term persistence of the volatility. Hsieh (1995) noted that nonlinear processes can generate this type of behavior while linear process cannot. Volatility clustering, also called ARCH effect (Engle, 1982) is a clear manifestation of the existence of non-linear dependences between returns observed at different lags. (FI)-GARCH (Baillie et al., 1996), $\alpha$-ARCH (Diebolt and Guegan, 1991), multifractal[1] (Barral and Mandelbrot, 2002; Mandelbrot et al., 1997; Lux, 2004) or many other stochastic volatility models (Heston, 1993; Taylor, 1994) have been used to describe the nonlinear behavior associated with volatility clustering and long-term memory. Many of the stylized fact of monovariate financial returns can be captured with multiplicative nonlinear processes, of which multifractal stochastic volatility models constitute a prominent example, for instance in the form of the so-called the "multifractal random walk" (Bacry et al., 2001) and its generalizations (Pochard and Bouchaud, 2002; Bacry and Muzy, 2003). See also Muzy et al. (2001) for a general multifractal multivariate generalization.

---

[1] While fractal objects, processes or measures enjoy a global scale invariance property – i.e., look similar at any (time) scale - multifractals only enjoy this property locally, i.e., they can be conceived as a fractal superposition of infinitely many local fractal models.

There is not doubt that volatility exhibits long-term dependence and can be predicted to some degree, see for instance (Harvey and Whaley, 1992; Alford and Boatsman, 1995; Fleming et al., 1995; Jorion, 1995; Figlewski, 1997; Sornette et al., 2003). Concerning the prediction of price changes or returns, the standard no-arbitrage argument states that the best predictor of future (discounted) prices is the current price. Interpreted with linear predictors, this implies the absence of linear time dependence. But nonlinear dependence can exist, leading to the possibility of developing nonlinear predictors that would violate the no-arbitrage argument. This could occur due to deviations from equilibrium (Fama, 1991), which could be taken advantage of by active investment strategies. Standard price equilibrium models, such as the CAPM (see Fama and French (2004) for a recent review) and APT (see Roll (1994) for a review), are based on the idea that the expected excess return of an asset is proportional to the expected covariance of the excess return of this asset with the excess return of the market portfolio, thus emphasizing the linear correlation between assets. Nonlinear extensions of the CAPM beyond the mean-variance approach have been proposed (Rubinstein, 1973; Krauss and Litzenberger, 1976; Harvey and Siddique, 2000; Fang and Lai (1997; Hwang and Satchell, 1999; Alexander and Baptista, 2002; Polimenis, 2002; Malevergne and Sornette, 2002), in order to account more accurately for the risk perceptions of investors, but they have not really improved the ability of the CAPM and its generalization to explain relative asset valuations. In a universe parallel to that of the academic community, the active management community is working on the premise that it pays to pursue active investment strategies based on the notion that there exists information available that has not been fully integrated into current market prices (Grossman and Stiglitz, 1980). This has led to studies suggesting some evidence for anomalous earnings supporting so-called technical analysis strategies (DeBondt and Thaler, 1985; Lo and KacKinlay, 1990; Jegadeesh and Titman, 1993; Bauer and Dahlquist, 1999; Andersen et al., 2000; Lo et al., 2000), going beyond the traditional approaches (serial correlation, run test, etc) which are themselves too unsophisticated or too restrictive to pick up complicated patterns of the price behavior.

Our contribution is to consider perhaps the simplest class of nonlinear models, which are compatible with the absence of linear predictability of price variations (or returns) but allow for some more complex nonlinear predictability. In contrast with technical analysis, the analytical specification of the model allows us to obtain a detailed mathematical and numerical understanding. The simplest nonlinear model we consider is the bilinear model, which is a special case of the general Volterra discrete series (Schetzen, 1980). It provides the most natural generalization of linear models. Bilinear models incorporate the class of linear models considered by Box and Jenkins (1976), namely the integrated auto-regressive moving average (ARIMA) models as special cases. In an unpublished empirical work, Lee (1981) reported that the best subset of bilinear models performs better than any kind of auto-regressive models, based on the Akaike's Information Criterion (AIC). Our goal is to develop an understanding of the underlying properties of such bilinear models and test further its potential predictive power.

In the sequel, we study the simplest example of this general class of Volterra discrete time series. We first address the question of how to estimate the structural parameter $b$ which measures the amplitude of the nonlinearity. We also discuss how to formulate the prediction of future values from past observations.

The paper is organized as follows. Section 2 introduces the model in terms of the dynamics of a random nonlinear variation *r(t)* and its generalizations in the class of Volterra discrete series. Some properties are derived, such as the distribution of *r(t)* shown to have an exponential tail, in agreement with empirical evidence on financial series at intermediate time scales. The response to an impulse is also calculated and is shown to give rise in some cases to explosive trajectories according to an exponential of an exponential growth. Section 3 focuses on the estimation of the nonlinear parameter *b* of the model, via the methods of moments. We discuss separately the questions of estimating the sign and the amplitude of *b*. Section 4 presents a scheme for the prediction of the next realization *r(t+1)* from the observation of the past time series. This section relies on two Appendices which explore different properties of the model. In particular, Appendix A teaches us some inherent instability of the inversion of the parameters of the process by testing the inversion method on a simplified linear moving average model. Appendix B presents the nonlinear lag-operator formalism which underlines the inherent instability in the estimation of the parameters in the regime of strong nonlinearity. Section 4 ends with a presentation of three metrics quantifying the properties of our prediction scheme. Section 5 concludes.

## 2. Presentation of the model and of its basic properties

We study the following bilinear process for the variable $r(t) = \ln[p(t)/p(t-1)]$ (the relative price variation or return) in discrete time steps:

(1) $\quad r(t) = e(t) + b\, e(t-1)\, e(t-2)$

where *e(t)* is a white noise Gaussian process $N(0,s)$ with zero mean and standard deviation (std) *s*. For *b=0,* the process (1) reduces to an i.i.d. white noise for a constant *s* or to more general stochastic volatility processes if *s* has itself a stochastic dynamics.

The process (1) is stationary (in a narrow sense) since any finite set of $r(t_1),\ldots,r(t_k)$ is defined through the *stationary* sequence of i.i.d. e(t)'s. If e(t) has a finite variance, then r(t) is besides a stationary process in a wide sense (of the second order).

We shall be mainly concerned with the predictability of the time series (1), that is, how well can *r(t+1)* be forecasted from the knowledge of the set of all past values *{r(t),r(t-1),r(t-2), …r(1), r(0)}*. This problem of predicting *r(t+1)* can be reduced to two separate problems:
   (i) the estimation of the parameters *s* and *b,* and
   (ii) the restoration of the innovations *e(u)* ,   $u = t - T,…, t$, from the observed realization *r(u)* ,   $u = t - T,…, t$. We call the latter problem *the inversion.*
These two problems are considered in turn below. We then discuss different implementations of prediction schemes and assess their performance.

### 2.1 Two-point and three-point correlation functions
A first characterization of the process (1) is through correlation functions. The process is centered, with *E[r(t)]=0,* and a zero correlation function *Corr[r(t),r(t')]=0* for t different from t'. The model (1) has thus zero linear correlation, which translates into an absence of linear predictability. However, the following three-point correlation function is non-zero and reads

(2)   $E[r(t-2)\ r(t-1)\ r(t)] = b\ s^3$.

And more generally, we have $E[r(t_1)\ r(t_2)\ r(t_3)] = b\ s^3\ \delta(t_3-t_1-2)\ \delta(t_2-t_1-1)$ for $t_3>t_2>t_1$. This implies that it should be possible to predict in part the next return $r(t)$ already from the knowledge of the two past returns $r(t-1)$ and $r(t-2)$. The process (1) has thus zero linear predictability and some non-vanishing nonlinear predictability, whose efficiency will be quantified below.

In practice as we shall see, the problem of predicting the future value $r(t+1)$ of the process (1) is very difficult. If the parameter $b$ is small, then $r(t)$ is close to the white noise $e(t)$, which is by definition unpredictable. If $|b| >> 1$, the nonlinear term dominate and $r(t+1) \cong b\ e(t)\ e(t-1)$ which, together with the stochastic innovations, leads to serious instabilities in the inversion of the innovations from the observed returns, a necessary step for the prediction.

From expression (1), we have $[r(t)]^2 = [e(t)]^2 + 2be(t)\ e(t-1)\ e(t-2) + b^2\ [e(t-1)]^2\ [e(t-2)]^2$. This shows that the covariance $Cov[[r(t)]^2[r(t+\tau)]^2]$ is zero for $|\tau|>2$ if the $e(t)$'s are white noise, as assumed here. While the volatility $[r(t)]^2$ has a two-step memory, thus larger than the zero-step memory of r(t) itself, this is not what one could refer to as "long memory". Thus, the process (1) with i.i.d. $e(t)$'s has the bad property of not describing one of the stylized fact of financial returns, namely the long memory of the volatility. This of course can be easily cured by putting the long memory of the volatility in the $e(t)$'s. Since we focus here our attention mostly on the predictability of $r(t)$, that is of its sign to a large degree, this question is secondary.

**2.2 Generalization to other bilinear and trilinear stochastic models**
It may be interesting to extend (1) to take into account some specific documented properties of the empirical returns, such at the "leverage effect" of a negative correlation between past returns and future volatility (Black, 1976; Christie, 1982; Figlewski and Wang, 2000; Bouchaud et al., 2001): $E[r(t)\ |r(t+1)|^2] < 0$. For such leverage effect to exist, the simplest specification is

(3)   $r(t) = b\ e(t-1)\ e(t-2)\ +\ c\ e(t-1)\ e(t-3)$ ,

which ensures that $r(t)\ |r(t+1)|^2$ contains a term proportional to $b^2\ c\ e(t)^2\ e(t-1)^2\ e(t-2)^2$ coming from the first term of $r(t)$ multiplied by the cross-product of the two terms of $r(t+1)$. This product of squares has a non-zero expectation with a sign controlled by that of $c$, which must thus be negative for the standard leverage effect to hold. The process (4) is such that the reverse time ordering r(t+1) $|r(t)|^2$ has zero expectation, which is requested as the volatility at time t does not predict return at time t+1. One would also have to check that other terms do not give non-zero contributions, i.e., products of even-order powers of the e's for r(t+1) $|r(t)|^2$).

Another characteristic structure of empirical returns is the skewness $E[r(t)^3]/(std(r(t))^3$ which is often found negative for empirical returns. It can be constructed from terms like

(4)   $r(t) = e(t) + b\ e(t-1)\ e(t-2)\ +\ g\ e(t)\ e(t-1)\ e(t-2)\ + ...$

which yields

(5)   $E[r(t)^3] = b\ g\ E[\{e(t)\ e(t-1)\ e(t-2)\}^2] = b\ g\ s^6$,

and thus *b g* must be negative. It is easy to check that the excess kurtosis is non-zero for (1), (3) and (4) (see below), showing that such multilinear models produce tails fatter than the Gaussian law.

**2.3 Underpinning of bilinear stochastic models in Volterra discrete series**
These above models belong to so-called Volterra discrete series (Schetzen, 1980). The Volterra series constitute very general means of describing a continuous-time output $y(t)$ in terms of an input $x(t)$. The Volterra series expansion for a causal, time-invariant system can be expressed as

(6)     $x(t) = H_1[e(t)] + H_2[e(t)] + H_3[e(t)] + ... + H_n[e(t)]$

in which the n-th degree Volterra operator $H[.]$ is defined by the convolution

(7)     $H_n[e(t)] = \int_0^{+\infty} ... \int_0^{+\infty} h_n(t_1, .., t_n) \, e(t-t_1) \, ... \, e(t-t_n) \, dt_1...dt_n$

and the Volterra kernels $h_n(.)$ have unspecified forms but $h_n(t_1, .., t_n)=0$ for any $t_i<0$, $i=1, 2, ..., n$. In discrete time, this equation becomes (Fakhouri, 1980)

(8)     $H_n[e_{t\Sigma}] = \Sigma_{j_1=0}^{+\infty} ... \Sigma_{j_n=0}^{+\infty} \, h_n(j_1, .., j_n) \, e_{t-j_1} \, ... \, e_{t-j_n}$

Expression (6) with (8) provides a natural generalization of linear system theory: for a linear system, $x(t) = H_1[e(t)]$, the first degree kernel $h_1(t)$ is the impulse response, which completely describes the system. For higher-degree systems, $h_n(t_1, .., t_n)$ can be thought of as a n-dimensional impulse response. Volterra series expansions model the output $x(t)$ of a system as a polynomial in the delayed inputs $e(t)$, $e(t-1)$, $e(t-2)$, … As an example of application, multivariate nonlinearity tests have been applied by Harvill and Ray (1998) to a set of seasonally adjusted quarterly capital expenditure and appropriations in U.S. manufacturing to construct bilinear models of the form (6) with (8).

We should note however that the terminology "bilinear" is often used to refer to different processes of the form (Rao, 1981)

(9)     $x(t) = \Sigma_{i=1}^{p} a_i \, x(t-i) + \Sigma_{i=0}^{q} c_i \, e(t-i) + \Sigma_{k=1}^{m} \Sigma_{l=1}^{n} b_{k,l} \, x(t-k) \, e(t-l)$

where $\{e(t)\}$ are independent and identically distributed (i.i.d.) random variables. Under certain conditions, such bilinear processes can be written in a Markovian representation of the form $E(t) = A(t) \, E(t-1) + B(t)$, $X(t) = H \, E(t-1) + G(t)$, where $\{A(t), B(t)\}$ is an i.i.d. sequence of random matrices depending on the model parameters (Turkman and Amaral Turkman, 1997). This makes these bilinear processes part of discrete stochastic recurrence equations (Kesten, 1973; Le Page, 1983; Goldie, 1991), also known in Physics as linear discrete delay equations with multiplicative noise. Least-square estimators of the parameters of these bilinear models (Grahn, 1995; Bibi and Oyet, 2004) and conditional maximum likelihood estimation procedures (Dai and Billard, 2003) have been developed.

**2.4 Probability density function (PDF) of the sequence of random variables (rv) generated by the bilinear model (1)**
We now derive the PDF of the set $\{r(t)$, $t=0$ to $T$, with $T \to +\infty$, where $r(t)$ is the process defined by expression (1). Consider the random variable (rv) $\eta = \xi_1 + b \, \xi_2 \, \xi_3$, where $b$ is a fixed parameter and $\xi_1$, $\xi_2$, $\xi_3$ are standard i.i.d. Gaussian rv with PDF $g(x) = (2\pi)^{-1/2} \exp(-x^2/2)$. The conditional density of $\eta$ under fixed $\xi_2 = y$ is Gaussian density with zero mean and variance $1+b^2y^2$. Averaging it over $y$, we get the unconditional density of $\eta$:

$$(10) \quad f_\eta(x) = (2\pi)^{-1/2} \int_{-\infty}^{\infty} \exp(-y^2/2) \, (2\pi(1+b^2y^2))^{-1/2} \exp(-x^2/(2(1+b^2y^2))) \, dy =$$

$$= (1/b\pi) \int_0^{\infty} \exp(-\sinh^2(z)/b^2 - x^2/\cosh^2(z)) \, dz$$

$$(11) \quad f_\eta(x) = (1/b\pi)\exp(b^{-2}) \int_0^1 (1/(1-w^2)) \exp(-x^2(1-w^2) - 1/(b^2(1-w^2))) \, dw.$$

Unfortunately, these integrals cannot be expressed with known special functions, but expression (11) is suitable for numerical integration leading to a numerical evaluation of density $f_\eta(x)$.

A saddle-point estimation of the integral in (11) shows that $f_\eta(x) \sim bx \exp[-2|x|/b]$ is a symmetrical function decreasing exponentially for large $|x|$ (typically the exponential holds for $|x|>b$). Intuitively, the Gaussian PDFs of the $\xi_i$'s are transformed into an exponential PDF by the fact that large values of $f_\eta(x)$ occur typically when both $\xi_2$ and $\xi_3$ are large and of the same order of magnitude $\xi_2 \sim \xi_3 \sim \sqrt{x}$, which occurs with the probability proportional to $g(\sqrt{x})g(\sqrt{x})$ which is exponential in $x$. The above derivation extends straightforwardly for the tail of the PDF of the rv $\eta = \xi_1 + \xi_2(b_3\xi_3+...+b_m\xi_m)$, since the rv $b_3\xi_3+...+b_m\xi_m$ is independent of $\xi_1$ and $\xi_2$, has zero mean and variance $b_3^2+...+b_m^2$.

As a consequence, for the process (1) in which the $e(t)$'s are white noise Gaussian rv with distribution $N(0,s)$ (with zero mean and standard deviation $s$), the PDF of the rv $r(t)$'s is symmetric with two exponential tails proportional to $\exp(-2|x|/bs^2)$. If $b<0$, the same result holds with $b$ replaced by $|b|$. The exponential law has been found to describe well the distribution of financial returns at intermediate time scales, from hours to weeks, either in the tail (Mantegna and Stanley, 1995; Cont et al., 1997) and even over the full range (Laherrère and Sornette, 1998; Dragulescu. and Yakovenko, 2002 ; Silva and Yakovenko, 2003 ; Silva et al., 2004). Thus, model (1) provides a reasonable representation of one of the basic stylized fact that the empirical distributions of returns have tails heavier than a Gaussian and can be approximated by the exponential law.

Note that the PDF of the rv such as $\eta = \xi_1 + b\,\xi_2\xi_3\xi_4$, which involves triplet products can be calculated similarly. As an illustration, the characteristic function $\varphi(t)$ of this rv $\eta = \xi_1 + b\,\xi_2\xi_3\xi_4$ is

$$(12) \quad \varphi(t) = \exp(-t^2/2) \, (2\pi)^{-1/2} (1/t) \exp(1/4t^2) K_0(1/4t^2),$$

where $K_0(x)$ is the modified Bessel function of zero order, so that the tail of the PDF of $\eta$ is of the form $\exp(-a\eta^{2/3})$ where $a$ is a constant (Frisch and Sornette, 1997), providing a simple mechanism for stretched exponential distributions of returns (Laherrère and Sornette, 1998; Malevergne et al., 2005).

## 2.5 The response of the inversion of the nonlinear model to an impulse: condition for stability of the inversion of the innovations *e(t)* from the observed *r(t)*

This section derives analytically a sufficient condition for the *stability* of the nonlinear inversion which derives *e(t), e(t-1), ...* from *r(t), r(t-1), r(t-2), ...* For this, we rewrite (1) in a form that constructs the current innovation *e(u)* through the current process value *r(u)* and two previously determined innovations *e(u-1)* and *e(u-2)*:

(13)   $e(u) = r(u) + b' e(u-1) e(u-2)$ ;   $u = 1, 2, ...,$   where $b' = -b$.

We may consider the relation (13) as the *nonlinear response* of the output innovations to the input process *r(u)*. In order to investigate the stability of this inversion, let us consider a *special impulse input*:

(14)   $r(u) = \begin{cases} a; & a > 0; \quad u = 1, 2; \\ 0; & u \neq 1, 2, \end{cases}$

and we assume $e_0 = e_{-1} = 0$. Due to the structure of the nonlinear model (1), the impulse must last at least two time steps to produce a non-zero response. It should also be noted that the general theory of response functions of linear systems cannot be applied to the nonlinear system (1) without special studies of the relevant effects. As the general known properties of linear systems cannot be expected to apply directly to (1), we shall prove all necessary assertions for the NL model (1).

Then, iterating (13), we get the following response:

(15)   $e(1) = a;\quad e(2) = a;\quad e(3) = b' a^2;\quad e(4) = b'^2 a^3;\quad e(5) = b'^4 a^5; \; ....$

Focusing on the stability of the inversion, we restrict our attention to the amplitudes of the *e(u)*'s. We thus take the modulus of the *e(u)*'s and go to logarithms, denoting

(16)   $v_u = \log |e(u)|$ ;   $\gamma = \log(|b|);$   $\mu = \log(a)$.

This yields the following difference equations for $v_u$:

(17)   $v_1 = \mu$ ;   $v_2 = \mu$ ;   $v_3 = \gamma + 2\mu$ ;   $v_4 = 2\gamma + 3\mu$ ;   $v_5 = 4\gamma + 5\mu$ ; ...

We can write the general term of the sequence (17) in the form:

(18)   $v_k = \Gamma(k)\gamma + M(k)\mu$,

where $\Gamma(k), M(k)$ are some functions of *k* to be determined. We obtain the following relations connecting these functions with three sequential arguments:

(19)   $\Gamma(k) = \Gamma(k-1) + \Gamma(k-2) + 1;$   $M(k) = M(k-1) + M(k-2)$.

Taking account of the initial values of $\Gamma(k), M(k)$ determined from (17), namely $\Gamma(0) = \Gamma(1) = 0$ and $M(0) = M(1) = 1$, we obtain the following solutions of the difference equations (19):

(20)     $\Gamma(k) = (1/\sqrt{5})[(1+\sqrt{5})/2]^k - (1/\sqrt{5})[(1-\sqrt{5})/2]^k$ ;  $M(k) = \Gamma(k) + 1$;   $k = 1, 2, ...$

The numbers *M(k)* are well-known in mathematics as *the Fibonacci numbers,* and the first equation in (20) is called *the Bine formula.* Note that this approach in terms of the difference equations permits to determine the power indices in more general cases. Returning to the amplitudes $|e_u|$'s, expression (16) with (18) gives

(21)    $|e(u)| = |b|^{\Gamma(u)} a^{\Gamma(u)+1} = a \, (|b| \, a)^{\Gamma(u)}$ .

Thus, if  $|b| \, a > 1$ , then $|e(u)|$ explodes to infinity super-exponentially (specifically, as the exponential of the exponential of time), corresponding to an unstable inversion. For $|b| \, a < 1$, $|e(u)| \to 0$ and the inversion is stable. One can conclude that, if in a realization of the process *r(u)*, two large values (of absolute amplitude *a*) occur such that the inequality  $|b| \, a > 1$ is verified, then a divergence of the recurrent procedure is quite possible. Of course, there is a chance that this "burst" will be compensated by nonlinear effects caused by neighboring values of the process $r_u$ , but there is still some non-zero probability for the burst to persist. Thus, one could take the conditions

(22)                           $|b| \, a < 1$

as *a necessary condition* for an unstable inversion (here *a* is a large amplitude of two sequential values of the process  *r(u)*). Knowing *b,* one could, generally speaking, calculate the probability of such an event for a stationary process with known distribution. One important conclusion can now be made: if the distribution of the process *r(u)* is unlimited, then the stability of the inversion for the NL-model (1) is fulfilled only with *some probability less than one* (although this probability can be very close to one). Thus, the probability for a stable inversion depends *on the length of the realization.* This conclusion will be further illustrated by our attempts in section 4 to perform inversions and predictions.

**3. Moment estimation of the nonlinear parameter *b***

We now turn to the problem of estimating the parameter *b*, which is the first pre-requisite before developing prediction schemes for the series (1). Determining *b* is crucial since it contains both the information on the nonlinear structure and on the dependence of successive values of the time series *{r(t)}*, as seen from expression (2).

We cannot use the expression of the PDF of the variable *r(t)* derived in (11) to obtain a maximum likelihood estimation (MLE) of *b* because successive values of *r(t)* are not independent and thus the likelihood of *b* can not be written as the product of the PDFs (11) expressed for each *r(t)*.  In order to use the ML approach, we would need to derive the multivariate distribution of the set *{r(t)}*. In view of the complexity of the MLE, we turn to the method of moments.

**3.1 Definition and expressions of moments**
It is convenient to re-parameterize the system (1) by writing *e(t) = s e'(t),* where *s* is the standard deviation of the *e(t)*'s, to obtain random variables *e'(t)* which have unit variance. After this change of variable, we obtain the nonlinear (NL) equation *r(t) = s [e'(t) + bs e'(t-1) e'(t-2)]*. We rewrite it using the correspondence e' $\to$ e  and  bs $\to$ b:

$$(23) \quad r(t) = s\,(e(t) + b\,e(t-1)\,e(t-2)), \qquad t = 0, \pm 1, \pm 2,\ldots$$

where $e(t)$ is now a standard Gaussian white noise with unit std, and $b$ is a dimensionless parameter. The parameters $(s,b)$ are unknown. We propose to use for the estimation of these parameters all non-zero moments of the 2-nd, 3-rd and 4-th order as well as the absolute moment of the first order. These moments are:

$$(24) \quad E[|r(t)|] = (s/(2\pi b))\exp(1/4b^2)[K_0(1/4b^2) + K_1(1/4b^2)];$$

$$(25) \quad E[r(t)^2] = s^2\,(1 + b^2);$$

$$(26) \quad E[r(t)\,r(t-1)\,r(t-2)] = s^3\,b;$$

$$(27) \quad E[r(t)^4] = 3\,s^4\,(1 + 2b^2 + 3b^4);$$

$$(28) \quad E[r(t)^2\,r(t-1)^2] = s^4\,(1 + 4b^2 + 3b^4);$$

$$(29) \quad E[r(t)^2\,r(t-2)^2] = s^4\,(1 + 4b^2 + b^4).$$

In (24), $K_0(z)$, $K_1(z)$ are the modified Bessel functions of the zero and first orders respectively. The moment estimation method consists in calculating the sample analogs of these moments, equating them to the theoretical expressions (24-29) and solving for $(s,b)$. This gives six equations (24-29) in which the theoretical expectation $E[\,.\,]$ in the l.h.s. is replaced by the sample averaging operation $<\ldots>$. We shall denote the sample values of the process $r(t)$ by $z_t$.

It is important to stress that, among the moments $E[|r(t)|]$, $E[r(t)^2]$, $E[r(t)\,r(t-1)\,r(t-2)]$, $E[r(t)^4]$, $E[r(t)^2\,r(t-1)^2]$ and $E[r(t)^2\,r(t-2)^2]$ calculated in (24-28), only $E[r(t)^2]$ and $E[r(t)\,r(t-1)\,r(t-2)]$ are function solely of the variance $E[e(t)^2]$ of the innovations. The other moments are in addition function of higher-order moments of the distribution of the $e(t)$'s. Thus, from an estimation view point, working only with $E[r(t)^2]$ and $E[r(t)\,r(t-1)\,r(t-2)]$ can be expected to give more robust and reliable estimations, less spoiled by the specific form of the distribution of the innovations. In other words, working only with $E[r(t)^2]$ and $E[r(t)\,r(t-1)\,r(t-2)]$ should give results not (or at least less) sensitive to the deviations from normality of the innovations.

It is easy to eliminate the parameter $s$ from these relations by dividing equations (24),(26,27) by equation (25) at the corresponding power. We thus obtain five equations containing only the unknown parameter $b$:

$$(30) \quad ((1+b^2)^{-1/2}/(2\pi b))\exp(1/4b^2)[K_0(1/4b^2) + K_1(1/4b^2)] = <|z_t|>/[<z^2_t>]^{1/2};$$

$$(31) \quad b/(1+b^2)^{3/2} = <z_t\,z_{t-1}\,z_{t-2}>/(<z^2_t>)^{3/2};$$

$$(32) \quad (1 + 2b^2 + 3b^4)/(1+b^2)^2 = (1/3)<z^4_t>/(<z^2_t>)^2;$$

$$(33) \quad (1 + 4b^2 + 3b^4)/(1+b^2)^2 = <z^2_t\,z^2_{t-1}>/(<z^2_t>)^2;$$

$$(34) \quad (1 + 4b^2 + b^4)/(1+b^2)^2 = <z^2_t\,z^2_{t-2}>/(<z^2_t>)^2.$$

Thus, we have 5 different equations for the estimation of *b*. We can use them in several ways, but we prefer the moment method based on equation (31) since as we said above, unlike the other equations (30), (32)-(34), it does not use any assumption on the Gaussian nature of the distribution of innovations *e(t)*. We consider this fact as an advantage of the method based on equation (31). We verify that the estimation of b thus obtained from (31) is more efficient, confirming the validity of the general reasoning.

The information on the sign of *b* is contained solely in the third order moment (26) (which becomes (31) after normalization). The determination of *b* can thus be decomposed into two tasks consisting respectively in the determination of (i) the sign of *b* and (ii) its amplitude |*b*|. We address these two tasks in turn. With no loss of generality and for simplicity of notation, we take s=1 which allows us to directly work with the non-normalized moments.

**3.2 Preliminary analysis of the statistical properties of the third-order moment**
The third-order moment *E[r(t) r(t-1) r(t-2)]* plays a special role because it is the first non-vanishing moment embodying some information on the dependence structure of the time series *{r(t)}*. As shown in expressions (2) and (26), the third-order moment is directly proportional to *b*, and thus vanishes for $b \to 0$, in contrast with all the other moments which have non-zero limits. The third-order moment can thus be expected to be the most pertinent source of information on the time series and on the determination of *b*. In addition, the estimates of *b* based on the third moment $< z_t \ z_{t-1} z_{t-2} >$ are in general more stable than estimates based on the $4^{th}$-order moments.

In this section, we investigate some properties of the third-order moment, of its sign and how best to determine the sign of the parameter *b* based on a nonlinear transform of the observed values $z_t$. This method turns out to be more efficient than the sample moment estimator *sign*( $< z_t z_{t-1} z_{t-2} >$ ) at least for |*b*|>1.

Fig. 1 shows the normalized third moment as a function of the parameter *b* for positive values; for negative values the graph is anti-symmetrical (the third moment is an odd function of *b*). This curve suggests that recovering the value of the parameter *b* will be feasible only in a range of *b*-values close to the maximum, that is, in a limited range around *b=1*.

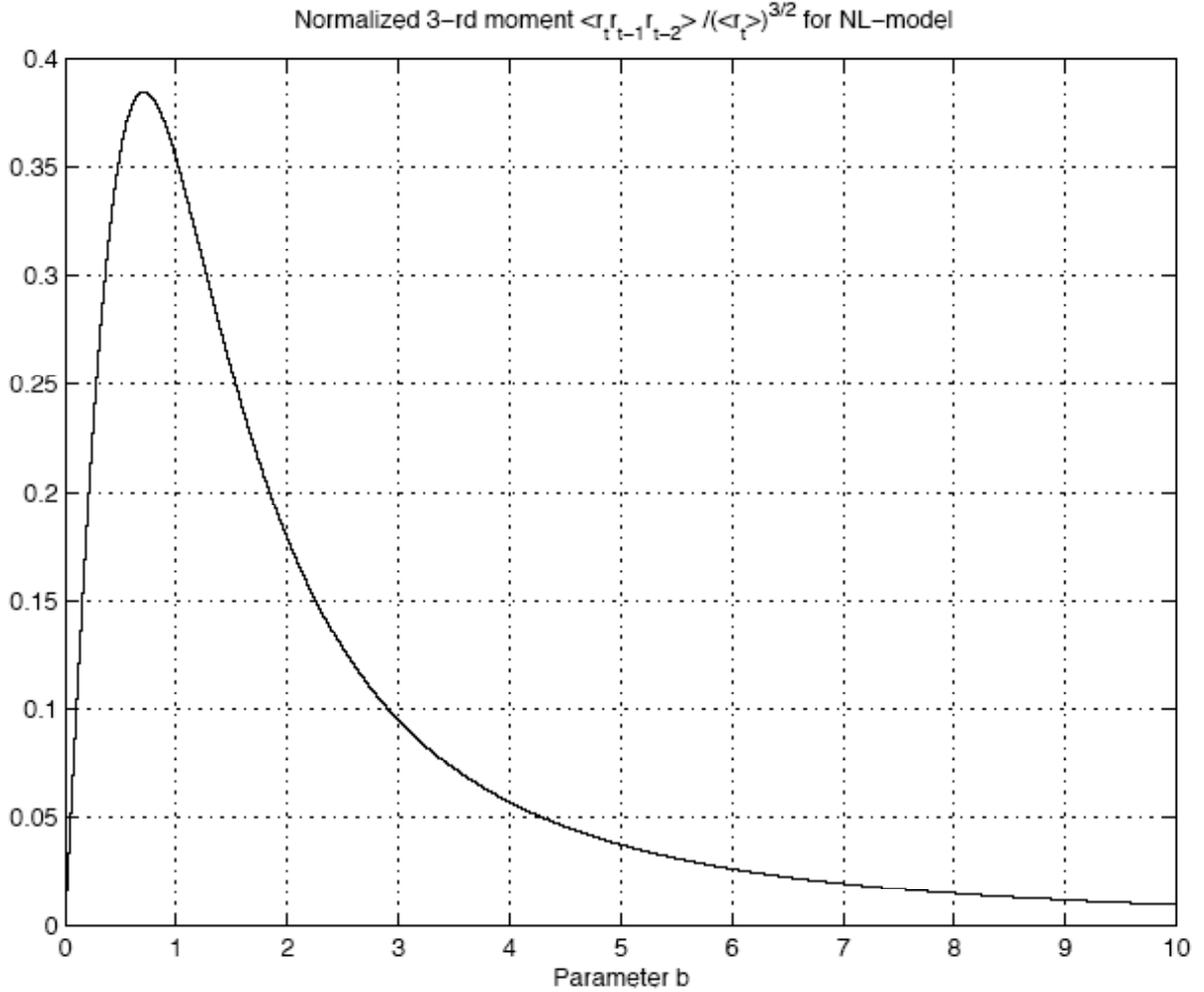

Fig. 1: Normalized third moment $E[r(t)\ r(t-1)\ r(t-2)]/E[r(t)^2]^{3/2}$ given by (31) as a function of the parameter $b$.

In order to understand the information contained in the empirical third-order moment, which is necessarily influenced by the limited statistics due to finite sample size, it is useful to study the distribution of the rv product $V = r(t)\ r(t-1)\ r(t-2)$. The analytical expression of the distribution of $V$ is very cumbersome (although it is possible to express it in an integral form). We find it more convenient to study the first four moments of this rv as functions of $b$. The first moment of $V$, which is nothing but $E[r(t)\ r(t-1)\ r(t-2)]$ is equal to $b$. The variance, skewness and kurtosis of $V$ are the following (we assume $s=1$ to simplify the notations):

$Var(V) = 1 + 12b^2 + 21b^4 + 9b^6$;

$Skewness(V) = (24b + 350b^3 + 1260b^5 + 1188b^7)/(1 + 12b^2 + 21b^4 + 9b^6)^{3/2}$ ;

$Kurtosis(V) = (27 + 492b^2 + 14790b^4 + 126054b^6 + 335799b^8 + 344250b^{10} + 99225b^{12})/$
$/(1 + 12b^2 + 21b^4 + 9b^6)^2$ .

These characteristics together with the coefficient of variation (std/mean value) are shown in Fig. 2a-2c and 3. The behavior of its kurtosis indicates that the distribution of V has a heavy tail. Indeed, the tail of the PDF of V is a stretched (sub-exponential) proportional to *exp[-a* $V^{2/3}$*]* (Frisch and Sornette, 1997). Notice also that the moments of *V* of even order grow fast with *b*. Thus, the expectation of *V* equals to *b* mainly because of rare large values of *V*, occurring with small probability.

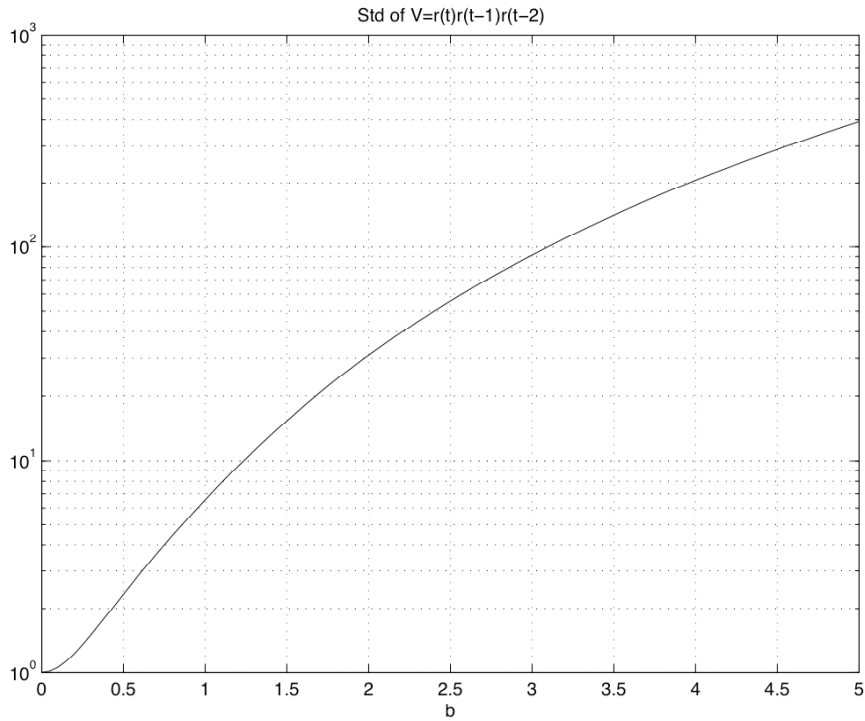

Fig.2a: Standard deviation of the rv *V=r(t) r(t-1) r(t-2)* for the system *r(t) = e(t) - b e(t-1) e(t-2)* with sdt *s=1* for the Gaussian innovations *e(t)* as a function of *b*.

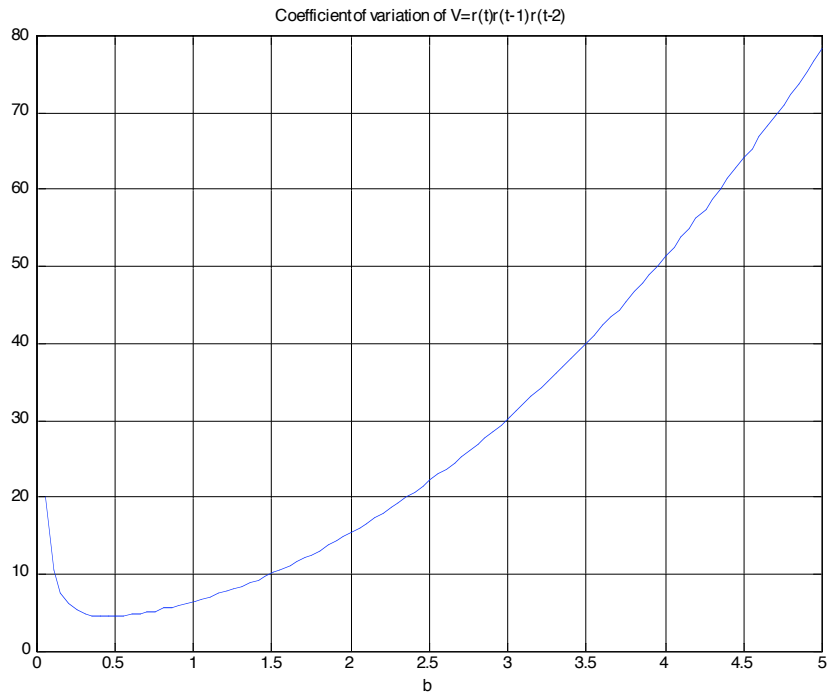

Fig. 2b: Coefficient of variation (std/mean value) of the rv *V=r(t) r(t-1) r(t-2)* for the system *r(t) = e(t) - b e(t-1) e(t-2)* with sdt *s=1* for the Gaussian innovations *e(t)* as a function of *b*.

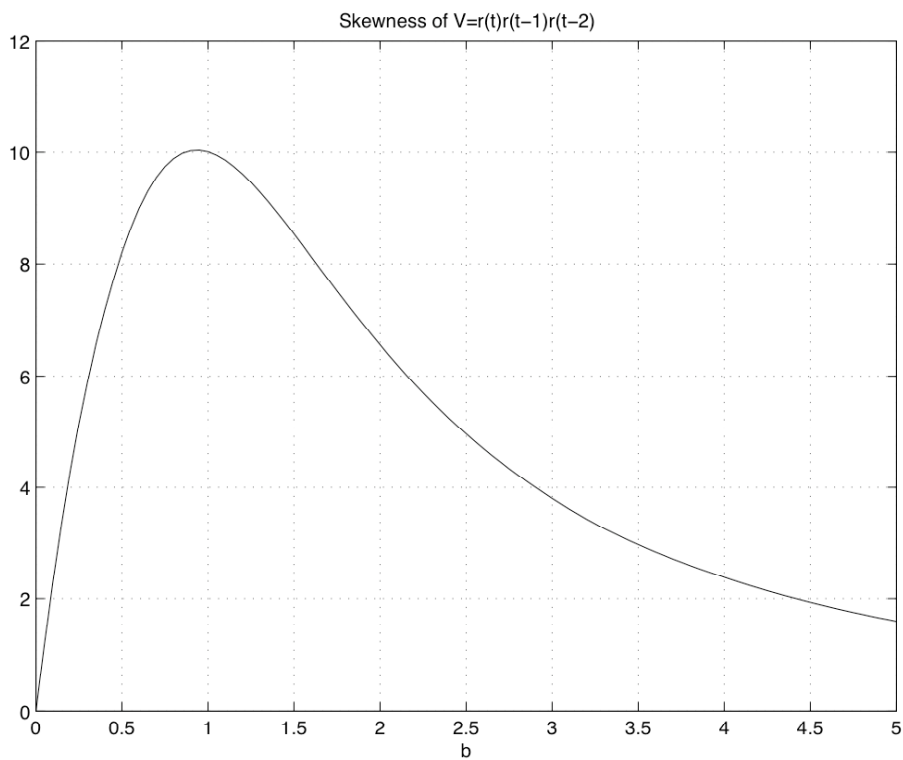

Fig. 2c: Skewness of the rv $V=r(t)\,r(t-1)\,r(t-2)$ for the system $r(t) = e(t) - b\,e(t-1)\,e(t-2)$ with sdt $s=1$ for the Gaussian innovations $e(t)$ as a function of $b$. The skewness for negative $b$ is just the opposite value.

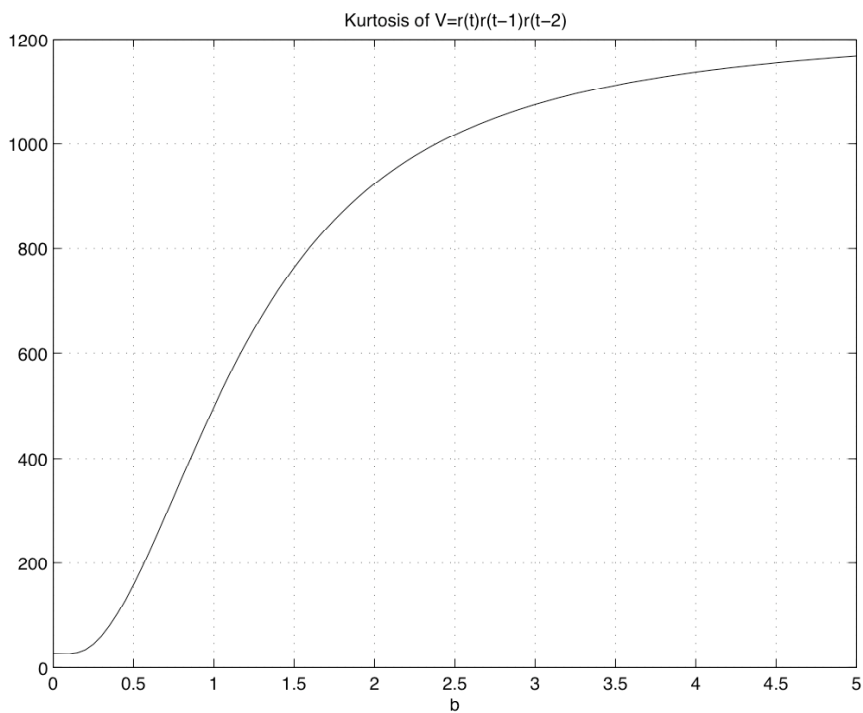

Fig. 3: Kurtosis of the rv $V=r(t)\,r(t-1)\,r(t-2)$ for the system (1), (13) with $s=1$ as a function of $b$ (in contrast to the skewness, the sign of $b$ has no influence on the kurtosis).

Fig. 2c shows that, for moderate *b*-values, the distribution of *V* is highly skewed in the direction of *sign(b)*, for the system *r(t) = e(t) + b e(t-1) e(t-2)*. Fig. 3 shows that the kurtosis is very large, which together with the large skewness confirms that the distribution of *V* is a heavy-tailed distribution. The coefficient of variation *std[V]/E[V]* of *V* is quite large, as its minimum is *4.64* for *b = 0.45* and increases fast for large values of *b*, again showing that there are large intrinsic fluctuations in the triple product *V*. In the presence of such large skewness and kurtosis, it is often useful to use the logarithmic transform preserving the information on signs, *W=sign(V)·log(|V|)*. The skewness of the PDF of *W* is of the same sign as *b* and ranges from *0* to a maximum of *0.11* for *b* close to *0.5*. The Kurtosis of the PDF of *W* peaks at *4.2* for b close to *0.5* and then decreases slowly for larger b's (for instance the kurtosis reaches *2* for *b=10*). The PDF of *W* is thus found approximately symmetrical and not far from a Gaussian law.

### 3.3 Estimation of the sign of parameter b

As we already said, the information on the sign of b can be obtained from the sign of the mixed third moment *<r(t)r(t-1)r(t-2)>*. Expressions (2) and (26) show that, on average (and we stress here the term "on average"), the sign of the sample moment *<r(t)r(t-1)r(t-2)>* coincides with the sign of *b*. As the previous section 3.2 has shown, there is so much fluctuation and skewness in the distribution of the triple product *r(t)r(t-1)r(t-2)* that significant errors in the determination of the sign of *b* are observed for intermediate and large values of *b*, especially for samples of moderate size (say, *200-1000* observations) which are typical for financial and many other applications.

In order to stabilize the moment estimate of *sign(b),* we suggest a new nonlinear *b*-sign estimator that turns out to be more efficient for |b| >1 than *sign (<r(t)r(t-1)r(t-2)>)*. The new sign estimator *Λ* has the following form:

(35)
$$\Lambda = 1; \quad if\ SampleMed \{ exp[r(t)r(t-1)r(t-2)] \} < 1;$$

$$\Lambda = -1; \quad if\ SampleMed \{ exp[r(t)r(t-1)r(t-2)] \} > 1;$$

where *SampleMed* means the sample median of the time series. We perform a numerical test of the efficiency of this estimator using a moving time window of length *L*. The comparison of this estimator with the straightforward estimator *V = sign (<r(t)r(t-1)r(t-2)>)* is shown in Table 1. We estimated the signs in time windows of length *L = 100, 200, 1000,* that move through the sample of total size $5·10^5$.

Table 1. Percentage of correct estimation of sign(b) by *V = sign (<r(t)r(t-1)r(t-2)>)* and by Λ defined by (35).

| b | | 0.3 | 1.0 | 1.5 | 3.0 | 5.0 |
|---|---|---|---|---|---|---|
| L = 1000 | V | 100 | 100 | 99.80 | 84.75 | 68.09 |
| | Λ | 100 | 100 | 100 | 100 | 99.90 |
|       | V | 99.94 | 99.32 | 93.85 | 70.88 | 58.28 |
| L = 200 | Λ | 97.34 | 99.82 | 99.44 | 99.93 | 87.56 |
|       | V | 99.23 | 97.65 | 89.44 | 66.27 | 57.76 |
| L = 100 | Λ | 90.47 | 97.96 | 96.93 | 88.77 | 79.08 |

Table 1 shows that the estimator $\Lambda$ is more efficient than $V$ for $b \geq 1$. It should also be noted that the most efficient estimation of the sign of $b$ occurs when $b$ is not far from *1*. When $b$ is very small or very large, any estimator becomes inefficient: for $b \rightarrow 0$, any information on the sign of $b$ is lost; for $b \rightarrow \infty$ this information is lost again since, in the representation $r(t) = e(t) + b\, e(t-1)e(t-2)$, we have approximately $r(t) \cong b\, e(t-1)e(t-2)$. Thus, at each time, the sign of $r(t)$ is almost independent of the sign of preceding values $r(t-1)$, $r(t-2)$... due to the innovation $e(t-1)$ which is specific to $r(t)$. Thus, the best domain for the estimation of the sign of parameter $b$ is located near $|b| \cong 1$.

### 3.4 Estimation of the amplitude $|b|$ from the normalized moment (31)

We examine the *b*-estimator provided by the solution of the equation (31). A solution of the absolute value of (31) exists if the following condition is fulfilled:

(36) $$|<z_t\, z_{t-1} z_{t-2}>/(<z^2_t>)^{3/2}| \leq 2/\sqrt{27} \cong 0.38.$$

Under this condition, the absolute value of equation (31) has one negative solution (it is inadmissible since we consider a positive $|b|$) and two positive solutions that we denote as $q_1$, $q_2$ ($q_1 \leq q_2$). The corresponding estimators $B_{11}$, $B_{12}$ of the parameter $b$ could therefore be

(37) $$B_{11} = q_1 ; \quad B_{12} = q_2.$$

When condition (36) is not fulfilled, we complement (37) by the following boundary values:

(38) $$B_{11} = \begin{cases} q_1; & \text{if } |<z_t\, z_{t-1} z_{t-2}>/(<z^2_t>)^{3/2}| \leq 2/\sqrt{27}; \\ 2/\sqrt{27}; & \text{if } |<z_t\, z_{t-1} z_{t-2}>/(<z^2_t>)^{3/2}| > 2/\sqrt{27}; \end{cases}$$

(39) $$B_{12} = \begin{cases} q_2; & \text{if } |<z_t\, z_{t-1} z_{t-2}>/(<z^2_t>)^{3/2}| \leq 2/\sqrt{27}; \\ 2/\sqrt{27}; & \text{if } |<z_t\, z_{t-1} z_{t-2}>/(<r^2_t>)^{3/2}| > 2/\sqrt{27}. \end{cases}$$

In order to distinguish between the two estimators $B_{11}$, $B_{12}$, we suggest to use the sample kurtosis value $k = <z^4_t>/(<z^2_t>)^2$. The corresponding theoretical counterpart is equal to $\kappa = 3(1+ 2b^2 + 3b^4)/(1+ 2b^2 + b^4)$ (for Gaussian innovations). $\kappa$ increases monotonically from 3 at $b=0$ (corresponding to the Gauss distribution) up to 9 at $b \rightarrow \infty$. For the boundary value $b =$ $1/\sqrt{2}$ dividing the two branches of the cubic equation (31), the theoretical kurtosis $\kappa$ is equal to *3.67*. We thus suggest the following rule for choosing between the two roots:

(40)          $k < 3.67$;          accept estimate $B_{11}$;

(41)          $k \geq 3.67$;          accept estimate $B_{12}$.

Table 2 presents the results of the method of moments for the |b|-estimates using the normalized moment (31) for simulated time series of the NL-model (1) for different values of the parameter *b*. We have also tested the other moment estimators obtained from expressions (24), (27-29), and found them inferior to that from the third-order moment (26),(31) (not shown). We also used a generalized method of moments (GMM) (Hansen, 1982; Hall, 2005), which consists in finding the value b which minimizes the generalized distance of the four sampled normalized moments taken simultaneously (given in the right-hand-side of expressions (30-34)) to their theoretical expression (given in the left-hand-side of (24-29)). We found that this GMM estimator is not better than *B12*, probably as a result of the large intrinsic variability of the process (1). These results confirm our arguments that the use of *B12* seems the most justifiable as it requires the weakest assumption on the distribution of innovations. This turns out to be also associated with more efficiency.

| *b* | 0.1 | 0.2 | 0.3 | 0.5 | 1.0 | 2.5 | 5.0 |
|---|---|---|---|---|---|---|---|
| *n=10⁶* (sample size); *w=10000* (length of time window); *m=198* (number of time shifts) *r = 5000* (interval of shift of time window) | | | | | | | |
| $B_{11}$ | 86 // 12.5 | 100// 5.05 | 100// 3.09 | 100// 1.53 | 57 // 0.33 | 98 // 0.67 | 83 //4.22 |
| $B_{12}$ | 86 // 0.036 | 100// 0.030 | 100// 0.033 | 100// 0.060 | 57 // 0.23 | 98 // 2.11 | 83 //4.76 |
| *n=10⁶* (sample size); *w=1000* (length of time window); *m=1998* (number of time shifts) *r = 500* (interval of shift of time window) | | | | | | | |
| $B_{11}$ | 60 // 14.7 | 85 // 8.29 | 97 // 4.45 | 94 // 1.85 | 63 // 0.83 | 78 // 5.12 | 47 //6.01 |
| $B_{12}$ | 60 // 0.10 | 85 // 0.089 | 97 // 0.10 | 94 // 0.13 | 63 // 0.41 | 78 // 2.15 | 47 //4.69 |
| *n=10⁵* (sample size); *w=100* (length of time window); *m=1998* (number of time shifts) *r = 50* (interval of shift of time window) | | | | | | | |
| $B_{11}$ | 43 // 6.7 | 52 // 9.09 | 57 // 7.19 | 66 // 5.48 | 61 // 4.48 | 35 // 8.24 | 24 //13.2 |
| $B_{12}$ | 43 // 0.25 | 52 // 0.20 | 57 // 0.18 | 66 // 0.22 | 61 // 0.60 | 35 // 2.17 | 24 //4.65 |

Table 2. Results of the method of moments for the |b|-estimates using the normalized moment (31) in moving time windows of length *w=10000* (top), *1000* (middle) and *100* (bottom) for simulated time series of the NL-model (1) for different values of the parameter *b*. For each true b-value, and each simulation step-up, each box gives the percentage of windows when an estimator exists (number to the left of //) and the root-mean square of the errors of the corresponding estimator (number to the right of //).

## 4. Prediction of the next step realization of the bilinear model (1)

We recall that predicting *r(t+1)*, given the time series of the past *{r(t), r(t-1), …, r(1), r(0)}* can be decomposed into two separate problems:

(i) the estimation of the parameters *s* and *b*, which has been addressed in the preceding section (and which is further exemplified in Appendix A in a linear version) and

(ii) the restoration (or inversion) of the innovations *e(u)*, *u = t - T,…, t*, from the observed realization *r(u)*, *u = t - T,…, t*.

The second problem involves a nonlinear inversion. Indeed, let us assume that one of the methods of statistical estimation provides us with a reasonable estimation of *b*, then the best predictor for *r(t+1)* knowing *r(t), r(t-1), r(t-2), …*, assuming that *r(t+1)* follows the process (1), is *r(t+1) = b e(t) e(t-1)*, where *e(t)* and *e(t-1)* need to be reconstructed (inverted) from the observed data *r(t), r(t-1), r(t-2), … , r(0)*. This predictor is best in the sense of the $L^2$-norm (variance): the conditional expectation of *r(t+1)* under a given past *r(t), r(t-1),…* is the best predictor in the sense of the least possible variance of errors. We now address the task of reconstructing the innovations *e(t), e(t-1),…*

### 4.1 Lessons from a similar linear model
As a guideline to better address the problems (i) and (ii) for the nonlinear stochastic model (1), it is useful to consider a simpler linear version, which keeps some of the elements of the more complex nonlinear problem. Appendix A presents a rather complete treatment of the simple moving average model of the first order:

(42)    *Y(k)=e(k)-be(k-1)*,    *k=1,2,…*    $|b| \geq 1$,

where *e(k)* are iid rv (white noise innovations) with zero expectation $E[e(k)]\equiv 0$ and standard deviation $\sigma = (E[e^2(k)])^{1/2}$. For $|b| \geq 1$, the inversion consisting of retrieving the *e(k)*'s from the *Y(k)*'s is unstable. But due to the linearity of (42), it is possible via a duality approach to map (42) onto another model of the same structure but for which the parameter *b* is replaced by *1/b* < *1*, which has a stable inversion (Hamilton, 1994). Since the two models (with parameters *b* and *1/b*) are equivalent via the duality, the inversion can always be performed satisfactorily for any value of *b*. Here, however, our goal is to learn how to address the instability occurring for $|b| \geq 1$ in the nonlinear model, for which we do not have the luxury of the duality approach (only valid for linear models). Appendix A thus explores all three problems that we encounter with the nonlinear stochastic model (1), namely the estimation of the parameter *b*, the inversion of the innovations *e(k)* and the prediction of the next time step (without using the duality property) to learn lessons that we next apply to the nonlinear model (1).

### 4.2 Lessons from a nonlinear lag-operator and nonlinear auto-regression approaches
In the Appendix B, we develop a non-linear lag-operator formalism to the bilinear stochastic recurrence equation (1). If the innovations *e(t)* are bounded below some constant in absolute value, this approach allows one to develop an explicit reconstruction of the innovations from the observed of the rv *r(t), r(t-1), … r(0)*. When the innovations *e(t)* are not bounded, as for instance when they are *N(0,1)* standard normal rv's, the nonlinear lag-operator provides another evidence confirming the results of section 2.5 that there is finite probability that the reconstruction procedure of the innovations will not work due to the super-exponential explosion of the response of the inversion of the innovations to some large impulses.

**4.3 Prediction of the next time step of the bilinear model by the inversion method.**
We consider the bilinear model (1) where *e(t)* are iid standard Gaussian rv, and $|b| \geq 1$, corresponding to the challenging unstable situation. We observe the sample *Y(1),…,Y(n)* and wish to predict the future value *Y(n+1)*. We suppose first that both the parameter b and the initial values *e(0), e(-1)* are known. Then we can reconstruct innovations *e(k), k=1,…,n* from *e(k)=Y(k)-be(k-1)e(k-2); k=1,…,n*. Since at *t=n*, we do not know *e(n+1)*, the best prediction obtained from equation (1) would be:

(43) $\qquad \hat{Y}(n+1) = be(n)e(n-1).$

Therefore, the problem of the prediction of *Y(n+1)* is reduced to the appropriate estimation of the unknown quantities *b, e(0), e(-1)*. But the main difficulty of this approach lies in the instability of the inversion procedure: small deviations in the estimates of *b, e(0), e(-1)* can result in exponentially (or even exponential of exponential) growing errors as a function of the time step *n* in the reconstructed innovations *e(n-1), e(n)*. In Appendix A, we have documented a few examples of such an instability in the linear MA model with $|b| >1$. Thus, because of this inherent instability which can only be worsen in the bilinear model, we are forced to use *moderate sample sizes n*. If the goal was limited to the estimation of the parameter *b*, and if *n* is large, we can use a *segmentation method*, cutting the sample of size *n* into *r* samples of size *s (n=r·s),* estimating b separately on each subsample of a small size *s*, and finally averaging the obtained estimates. Such an approach has been used successfully in on a similar problem (Pisarenko and Sornette, 2004). But, for the prediction of *Y(n+1)*, the segmentation approach is inapplicable: we have to use only the last segment of admissible length *s*. We thus propose following procedure seems sensible: for the estimation of the parameter *b*, we use the full accessible sample of *r(t)* whatever large it would be, whereas for the inversion procedure we use samples of length not more than the 10-15 last values r(t),r(t-1),…,r(t-15).

The innovations *e(k)* obtained from *e(k)=Y(k)-be(k-1)e(k-2)* are functions of three variables: *b, e(0)=v, e(-1)=w*. Under fixed *v,w*, it is easy to obtain explicitly the *conditional density* of the sample *Y(1),…,Y(n)* as follows.

For *n=1* we have the following (Gaussian) conditional density of Y(1):

(44) $\quad p(y_1 | v,w) \propto exp\{ -(y_1-bvw)^2/2 \}.$

Moreover, we can express *e(1)* through *Y(1)* as follows:

(45) $\quad e(1)=Y(1) – bvw.$

For *n=2* we have the following (Gaussian) conditional density of *Y(1),Y(2):*

(46) $\quad p(y_1,y_2 |b,v,w) = p(y_2 | y_1,b,v,w)\ p(y_1 |b,v,w),$

where *p(y$_2$ | y$_1$,b,v,w)* is the conditional density of *Y(2)* under fixed *Y(1),b,e(-1),e(0)*. We have for *p(y$_2$ | y$_1$,b,v,w)* :

(47) $\quad p(y_2 | y_1,b,v,w) \propto exp\{ -(y_2-be(1)w)^2/2 \},$

where $e(1)$ can be expressed as a function of $Y(1)$ and of the preceding innovation $e(0)=w$; $e(-1)=v$ by $e(1)=Y(1)-be(0)e(-1)$.

In this way, we can determine iteratively the conditional density of $Y(1),…Y(k)$ through the conditional densities of $Y(1),…,Y(k-1)$ determined earlier. We thus obtain the (Gaussian) conditional density of $Y(1),Y(2),…,Y(k)$:

(48) $\quad p(y_1,y_2,…,y_k |b, v,w) = p(y_k|y_1,…,y_{k-1} | b,v,w)\ p(y_1,y_2,…,y_{k-1} | b,v,w) =$
$\quad\quad\quad = exp\{ -( y_k-be(k-1)e(k-2) )^2/2 \}\ p(y_1,y_2,…,y_{k-1} |b, v,w),$

where the conditional density $p(y_1,y_2,…,y_{k-1},|b,v,w)$ was determined earlier, as well as the values $e(k-1),e(k-2),…e(1)$. Collecting all factors we get for the conditional likelihood $L(b,v,w)$:

(49) $\quad L(b,v,w) = p(y_1,y_2,…,y_n | b,v,w) \propto \prod_{k=1}^{n} exp\{ -( y_k-be(k-1)e(k-2) )^2/2 \}.$

and its logarithm

(50) $\quad Log( L(b,v,w) ) = -1/2 \sum_{k=1}^{n} ( y_k-be(k-1)e(k-2) )^2 .$

The conditional log-likelihood is equal to the negative squared sum of residuals of the approximation of the sample values $y_k = Y(k)$ by the products $b·e(k-1)e(k-2)$, where all past innovations can be determined recursively. One could use the conditional log-likelihood (50) to derive the conditional estimates of the parameters $b,v,w$. However, the conditional likelihood (50) is a very unstable function of its arguments, in particular if $n$ is large. In practice, this instability results in a very irregular form of the log-likelihood (50): its global minimum is reached somewhere near the true values of the parameters b,v,w but, even in a small vicinity of its minimal point, this function is very-very large. No standard method of search for the minimum, based of assumptions of some "smoothness" of the function, is applicable here (e.g. the Newton-Raphson method and similar approaches). The only remaining approach is to perform a systematic search on a very fine 3D grid (the grid step should decrease as $1/b^n$ with the used sample size $n$, as it was shown for the linear MA model with $|b| >1$ in Appendix A), otherwise large errors will result. This introduces significant practical restrictions for the use of the grid search method. Let us denote the estimates corresponding to the maximum of the log-likelihood (55) as $\hat{b},\hat{v},\hat{w}$. We use these estimates in order to calculate $e(n), e(n-1)$ from $e(k)=Y(k)-be(k-1)e(k-2)$. Let us denote the resulting values as $\hat{e}_n, \hat{e}_{n-1}$. We then use the values $\hat{b}, \hat{e}_n, \hat{e}_{n-1}$ for prediction of $Y(n+1)$ given by eq. (43).

The quality of the obtained estimation and prediction is estimated via the following metrics. As is often used, a first metric is the standard deviation of the estimate, calculated in a series of simulations (we usually use m=1000 simulations). For the quality of the prediction of Y(n+1), one could use the ratio

(51) $\quad \rho = std[Y(n+1) - \hat{Y}(n+1)] / std[Y(n+1)]$.

But it turns out that this characteristic is not always satisfactory, because the distribution of prediction errors of $Y(n+1) - \hat{Y}(n+1)$ is irregular, i.e. far from a bell-like shape. Specifically, the distribution of prediction errors contains two components: one has a regular bell-like shape concentrated around zero, whereas the other one consists of rare very large outliers. Typical calculations for *b=2* for 1000 simulations give that most of the errors are concentrated around zero in the interval (-3.5; 3.5), whereas there are a few large outliers of amplitudes of several hundreds. As a consequence, the standard deviation is not an appropriate characteristic of the efficiency of prediction, because it is too much depending on the occurrence of rare outliers. To address this problem, which is not only one of choosing the suitable metric but mainly of the existence of rare but very large instabilities, we modify our prediction procedure as follows: each time before issuing a prediction, we measure its amplitude given by equation (48). If the value given by (43) is less than some threshold *H*, we take it as the prediction. *If not - we restrain ourselves from issuing prediction*. Thus, the prediction scheme is modified to acknowledge that we are not able to make prediction at all time steps due to the occurrence of very strong instabilities. We would wish that the fraction of such "refusals" would not be large, otherwise such a scheme would be quite inefficient. Or alternatively, it could reveal only transient "pockets of predictability", as has been proposed recently (Andersen and Sornette, 2005). The estimation of the frequency of refusals will be denoted as $\theta$ and constitutes another parameter characterizing the quality of the prediction scheme. This modification allows us to quantify the quality $\rho$ of the prediction given by (51) only on the subset which makes sense intuitively. In addition, we estimate the probability of correctly predicting the sign sign[Y(n+1)] by sign[be(n)e(n-1)] as

(52) $\qquad \pi = P\{ sign[Y(n+1)] = sign[be(n)e(n-1)] \}$.

We thus characterize the quality of our prediction scheme by reporting $\rho$, $\theta$ and $\pi$.

We have carried out some preliminary experiments to determine an appropriate threshold *H* for our further experiments. In Table 3, we present our results for *b=2; v=0.3; w=-0.3; n=20*. The grid steps are: *Δb= 0.05; Δv= 0.1; Δw= 0.1*. The search for the maximum of the log-likelihood (50) was performed in the parallelepiped (1.5÷2.5)×(0÷0.6)×(-0.6÷0). It is necessary to note that, in this experiment, we purposely took a grid *containing the true parameter values*. Below, we show some results for experiments where this assumption is violated.

Table 3. Characteristics of the prediction for different values of threshold *H*. The parameters are *b=2; v=0.3; w=-0.3; n=20*. The grid steps are: *Δb= 0.05; Δv= 0.1; Δw= 0.1*. $\theta$ is the fraction of times when there are no acceptable predictions.

| H | 1 | 1.5 | 2 | 3 | 5 |
|---|---|---|---|---|---|
| $\rho$ | 0.722 | 0.757 | 0.738 | 0.791 | 0.814 |
| $\theta$ | 0.32 | 0.24 | 0.14 | 0.10 | 0.03 |

We consider as appropriate the threshold *H=2* which provides a value for $\rho$ ($\rho$ = 0.738) close to the minimum (the best achievement), and at the same time which keeps the probability $\theta$ of refusal at a low level ($\theta$ = 0.14). In this example, we obtain $\pi$ =0.641 ± 0.017, i.e., the probability to correctly predict the direction of the next price move (interpreting the next realization as the return for the next time period) is 64%.

In our other investigations below, we fixed the threshold *H* at *H = 2* and we explore the effect of *a random shift of the parameters b,v,w with respect to the grid*. For this purpose, in each random simulation, we shifted these parameters by a Gaussian perturbation with a small amplitude μ. If μ = 0, then the grid *contains the true parameter values*. For very small *μ*'s, the algorithm still works but, above some amplitude for *μ*, the grid search method fails. We consider this noise amplitude as the "last admissible noise amplitude" and we report it in the last column of Table 4. All parameters of the bilinear model and the grid are the same as above in Table 3.

Table 4. *Characteristics of the prediction for random shifts of b, v, w, with respect to the grid.* for *b=2; v=0.3; w=-0.3*

| | n=20 | | |
|---|---|---|---|
| μ | 0 | $10^{-6}$ | $5 \cdot 10^{-6}$ |
| ρ | 0.74 | 0.82 | 0.85 |
| θ | 0.17 | 0.20 | 0.26 |
| b | 1.93 ± 0.19 | 1.92 ± 0.20 | 1.92 ± 0.19 |
| π | 0.68 | 0.63 | 0.59 |
| | n=30 | | |
| μ | 0 | $10^{-9}$ | $10^{-8}$ |
| ρ | 0.51 | 0.52 | 0.60 |
| θ | 0.19 | 0.24 | 0.32 |
| b | 1.97 ± 0.10 | 1.984 ± 0.083 | 1.980 ± 0.094 |
| π | 0.73 | 0.72 | 0.68 |
| | n=50 | | |
| μ | 0 | $10^{-13}$ | $10^{-12}$ |
| ρ | 0.46 | 0.55 | 0.56 |
| θ | 0.22 | 0.32 | 0.40 |
| b | 1.9995 ± 0.014 | 1.9988 ± 0.023 | 1.969 ± 0.039 |
| π | 0.79 | 0.73 | 0.69 |

## 6. Concluding remarks

We have studied some statistical properties of an apparently very simple nonlinear model. Despite its simplicity, this model exhibit some typical nonlinear effects enhanced and complicated by its stochastic (multiplicative) nature. Its ordinary correlation function coincides with the *δ*-function, whereas the correlation function of absolute values of observed process (or the correlation function of some fractional power of absolute values) differs from the *δ*-function which permits in principle some prediction of this absolute value (although rather poor). The 1D-distribution function has exponential tails, whereas innovations are Gaussian rv. For comparatively large values of the parameter *b*, this model can be considered as a disturbed product of pair of innovations:

(53)    $z_u \cong -s\, b\, e_{u-1}\, e_{u-2}, \quad u = t - T, ..., t.$

Such products exhibit sometimes large excursions when two large innovations come one after other. Such excursions may cause large fluctuations in the *b*-estimates, not speaking of the errors on the prediction of future values. Of course, the innovations $e_u$ are unobservable, but large excursions of products of successive innovations are reflected (approximately) by corresponding excursions of the observable products $z_u\, z_{u-1}$, which can be recorded and studied. If one wishes to generalize the nonlinear model with the aim to get heavier tails and a longer range for the dependence of absolute values, it is possible to consider the following model:

$$(54) \qquad z_u = e_u + b\, S_u \cdot [\prod_{k=1}^{m} |e_{u-k}|\,]^{\rho}; \qquad u = t - T, ..., t.$$

where *m* is the order of the model; *b* is a *coefficient;* $\rho$ is a *power index;* *S* is the *composite sign* of past innovations:

$$(55) \qquad S_u = \prod_{k=1}^{m} sign(e_{u-k}).$$

For *m=2;* $\rho=1$ then the model (59) coincides with the model (1) which has been studied here. Varying the parameters *m,* $\rho$, one can get different decays of the tails of 1D distribution of $z_u$, as well as different long range dependence of $|z_u|$, while the ordinary correlation function of $z_u$ remains $\delta$-correlation. For the model (54) a useful "conjugate" statistics based on observable values $z_u$ would be:

$$(56) \qquad W_u = V_u \cdot [\prod_{k=1}^{m} |z_{u-k}|\,]^{\rho}.$$

Studying statistics of such type, one may hope to identify the "bursts" of $z_u$ and their predictability.

Returning to our nonlinear model (1), we can state that, whereas the problem of the estimation of the parameter *b* is more or less satisfactorily solvable as has been shown in section 3, the problem of predicting the future values of $z_u$ is solved only with very low efficiency. In a sense, the bilinear model (1) can be called *unpredictable.* One can hardly hope to fit satisfactorily this model to real economical time series, and to get any acceptable prediction for the future. But this model should become very useful for studying typical nonlinear effects of statistical non-linear models. These properties are expected from real financial returns for which future returns are often considered unpredictable. In this sense, our present study has shown, with one explicit example, that almost complete absence of predictability does not prevent the existence of significant three-point and higher order correlations as well as realistically complex properties. The stochastic bilinear model provides an interesting alternative to stochastic volatility models or to GARCH and their generalization.

**Acknowledgments:** This work is dedicated to the memory of Aaron Miles. We are grateful to Yannick Malevergne for useful feedback on the manuscript.

**Appendix A: Analysis of an unstable linear moving average model (without using the duality approach)**

Let us consider the simple moving average model of the first order:

(A1)  $\quad Y(k)=e(k)-be(k-1), \quad k=1,2,... \quad |b| \geq 1,$

where $e(k)$ are iid rv (white noise innovations) with zero expectation $E[e(k)]=0$ and standard deviation $\sigma = (E[e^2(k)])^{1/2}$. For $|b| \geq 1$, the inversion consisting of retrieving the $e(k)$'s from the $Y(k)$'s is unstable. We pay specially attention to this linear model, since the study of its instability, which is simpler to analyze than for the bilinear model (1), will allow us to extract some helpful conclusions, which can be used in the non-linear situation. As a consequence, we refrain from using the duality method, which applies only to linear processes (Hamilton, 1994), and use instead methods which can be also applied to nonlinear processes. In order to concentrate on the general aspects of the instability that can be common to non-linear models, we impose two additional simplifying assumptions that $e(k)$ is Gaussian white noise with unit standard deviation $\sigma = 1$. We shall discuss later how one can relax these two assumptions.

### A-1 Three methods for the estimation of the parameter $b$.

The covariance function of the process (A1) is:

(A2)  $\quad B(j) = 1 + b^2; \quad\quad j=0;$
$\quad\quad\quad\quad = -b; \quad\quad\quad\quad j=\pm 1;$
$\quad\quad\quad\quad = 0; \quad\quad\quad\quad\quad\text{other } j.$

We compare three methods for estimating the parameter $b$: the Maximum Likelihood estimate $b_{MLE}$, the moment estimate $b_{moment}$, and an estimate provided by the inversion of time series (A1) denoted as $b_{inv}$.

The MLE-estimate is calculated numerically by inverting numerically the full covariance matrix of $Y(k)$ of order $n \times n$ where $n$ is the length of the time series under study. The moment estimate is obtained from the equation:

(A3)  $\quad\quad\quad b_{moment} = 1/(n-1) \sum_{k=2}^{n} Y(k)Y(k-1).$

The standard deviations of these two estimates are estimated on 10000 realizations of $Y(k)$. Both estimates have practically no systematic bias.

The third estimate of $b$ is based on the inversion of the time series (A1). The method proceeds as follows. From equation (A1), we have:

(A4)  $\quad\quad\quad e(k)=Y(k)+be(k-1); \quad k=1,...,n.$

In order to be able to use this equation to obtain $e(k)$ from the $Y(k)$'s, two values are needed: $b$ and $e(0)$ which are a priori unknown. Choosing two values $\beta$ and $w_0$ for these two parameters, we then use (A4) with $b$ and $e(0)$ equal to these two fixed values $\beta$ and $w_0$. The corresponding reconstructed innovations are denoted $w(k)=w(k \mid \beta, w_0; b, e(0))$, and are given ~~by~~ recursively by

(A5) $\qquad w(k)=Y(k)+\beta w(k-1); \quad k=1,\ldots,n; \quad w(0)=w_0.$

Using the recurrence equations (A1) and (A5), we get:

(A6) $\quad w(k)=e(k)+(\beta - b)e(k-1)+ \beta(\beta - b)e(k-1)+ \beta^2(\beta - b)e(k-2)+\ldots+ \beta^{k-2}(\beta - b)e(1)+$
$\qquad + \beta^{k-1}(\beta w_0 - be(0)).$

In equation (A6), both $b$ and $e(0)$ are unknown, whereas we are free to choose $\beta$ and $w_0$. By setting $w_0 = 0$; $\beta = 1$, we get:

(A7) $\quad w(k)=e(k)+(1 - b)e(k-1)+ (1 - b)e(k-1)+ (1 - b)e(k-2)+\ldots+ (1 - b)e(1) - be(0).$

Now, assuming for the moment that $e(0)$ is fixed, we get from (A7) an unbiased conditional estimate of $-b \cdot e(0)$ in the form of $T_1$, since all the innovations $e(k), e(k-1),\ldots$ have zero expectation:

(A8) $\qquad T_1 = (1/n) \sum_{k=1}^{n} w(k).$

This means that the statistic $T_1$ can be used for centering $w(k)$ for any value of $e(0)$. We can now introduce the statistic $T_2$ to derive an estimate of the parameter $b$:

(A9) $\qquad T_2 = 1/n \sum_{k=1}^{n} [w(k) - T_1]^2.$

It can be shown that expectation of $T_2$ is equal to:

(A10) $\qquad E\, T_2 = 1-1/n -(1-b)(1-1/n)+(1-b)^2 (n/6-1/(6n)).$

Equating $T_2$ to its expectation yields the estimate $b_{inv}$ as the solution of the equation

(A11) $\qquad T_2 = 1-1/n -(1-b)(1-1/n)+(1-b)^2 (n/6-1/(6n)).$

Equation (A11) has two roots but, if $|b| \geq 1$, which we assume, it is easy to discriminate between these two roots and select the root $b$ with the correct sign.

Table A-1 gives the Mean Square Errors of the three estimates of $b$ for two sizes $n=10$ and $n=100$ of the time series. One can observe that the moment estimate has a larger std than that of the MLE-estimate, which is expected since the MLE is the most efficient asymptotic estimator. A superior quality of $b_{inv}$ (smaller MSE) as compared with $b_{MLE}$ for $n=10$, $b = 1$ and $n=10$, $b = 2$ suggests significant small sample size effects of the latter preventing the realization of its asymptotical properties of efficiency.. One can observe a significant asymmetry of the MSE of the estimate $b_{inv}$ with respect to the sign of the parameter $b$. For large $n$ ($n \geq 100$) and for $b$ close to unity, the estimate $b_{inv}$ can provide a rather accurate evaluation of the unknown parameter $b$.

Table A1. Mean-square error (MSE) of the three estimates: $b_{MLE}$, $b_{moment}$, and $b_{inv}$ described in the text.

|  | n=10 | | | n=100 | | |
| --- | --- | --- | --- | --- | --- | --- |
|  | $b_{MLE}$ | $b_{moment}$ | $b_{inv}$ | $b_{MLE}$ | $b_{moment}$ | $b_{inv}$ |
| b = -2 | **0.61** | 2.0 | 1.32 | **0.13** | 0.60 | 1.18 |
| b = -1 | **0.45** | 0.86 | 0.95 | **0.035** | 0.26 | 0.79 |
| b = 1 | 0.41 | 0.86 | **0.28** | 0.040 | 0.26 | 0.047 |
| b = 2 | 0.65 | 2.0 | **0.49** | 0.14 | 0.60 | 0.38 |

### A-2 Prediction of *Y(n+1)*.

The prediction requires two steps: (i) the estimation of the product $b \cdot e(0)$ and of the parameter $b$ and (ii) the use of the recurrence relation (A1).

#### i) Estimation of the product $b \cdot e(0)$ and of the parameter *b*.

Equation (A6) implies that the reconstructed innovations $w(k)$ under fixed $\mu$ and $w_0=0$ have the following conditional variances:

(A12)      $E[(w(k) - \mu)^2 \mid e(0)] = 1 + (\beta-b)^2 + \beta^2(\beta-b)^2 + \ldots + \beta^{2k-4}(\beta-b)^2$ ;    $k= 2,\ldots,n$

where we denote $-b \cdot e(0) = \mu$. For $\beta > 1$, $E[(w(k) - \mu)^2 \mid e(0)]$ given in (A12) increase exponentially with $k$ except for $\beta = b$. Thus, we can expect that, even for intermediate $n$ (say, $n \cong 20$), the random values $w(k)$ take very large values for all $\beta$ except in a small neighborhood $\beta \cong b$. This behavior suggests the following strategy for estimating both the unknown product $\mu = -b \cdot e(0)$ and the parameter $b$. We take $w(n)$ given by equation (A6) and normalize it by $\beta^{n-2}$. From the expression of $w(n)/\beta^{n-2}$, we obtain its conditional expectation and variance:

(A13)                $E[w(n)/\beta^{n-2} \mid e(0)] = \beta(\beta w_0 - be(0))$;

(A14)       $Var[w(n)/\beta^{n-2} \mid e(0)] = (\beta - b)^2 \beta^2 (1-1/\beta^{2n-2})/(\beta^2-1) + 1/\beta^{2n-4}$.

If $1/\beta^{2n-4} \ll 1$, which we assume, and if $\beta=b$ and $w_0 = e(0)$, then the expectation (A13) is zero and the variance (A14) is very small. Thus, the values of $\beta$ and $w_0$ which minimize (A13) and (A14) provide estimates for $b$ and $e(0)$. In practice, we perform a 2D grid search to obtain the global minimum of $|w(n)/\beta^{n-2}|$. The grid step must be very small (of order $1/b^n$) in order not to miss the global minimum. Large values of $n$ (say, $n > 100$) are undesirable from this perspective. On the other hand, large values of $n$ increase the accuracy of the estimation of $e(0)$ and $b$. These contradictory requirements restrict the domain of $b$ values for which reliable estimates of the parameter b can be obtained (see the examples below).

In summary, we follow the following estimation procedure. For a given sample $Y(1),\ldots,Y(n)$, we calculate $|w(n)/\beta^{n-2}|$ using (A5) as a two-dimensional function of the

parameters $(\beta, w_0)$ on a 2D grid. We search the pair $(\hat{\beta}, \hat{w})$ providing the minimum of $|w(n)/\beta^{n-2}|$ on the grid and take this pair as the estimate of the unknown values $(b, e(0))$.

### ii) Prediction of $Y(n+1)$.

In order to predict the future value $Y(n+1)$ at the present time $n$, we use equation (A1) with $k=n+1$, and omit the innovation $e(n+1)$ which is unknown at time $n$. We replace the unknown parameter $b$ by its estimate $\hat{\beta}$, and use for the innovations $e(n)$ the reconstructed estimates $w(n)$ calculated with $\beta=\hat{\beta}$ and $w_0=\hat{w}$ in accordance with the recurrence relation (A5). This leads to the following predictor for $Y(n+1)$:

(A15) $\qquad\qquad\qquad \xi_{n+1} = -\hat{\beta} w(n).$

Occasionally, an exponentially divergent "burst" occurs in the inversion process. They result in very large values of $|w(n)|$. We propose in such cases to replace the predictor (A15) by the trivial prediction $\xi_{n+1}=0$.

We calculated the percentage, called REF, of such "refusals". We measure the quality of the predictor (A15) by the ratio $\rho = std(Y(n+1) - \xi_{n+1}) / std(Y(n+1))$: the smaller $\rho$ is, the better is the quality of the prediction. All simulations were repeated $m=1000$ times. The grid step was taken equal to $0.025$ both for $b$ and for $e(0)$. The grid covered the rectangle $(1 \leq b \leq 5; -2 \leq e(0) \leq +2)$. As we noted above, using a grid with a finite step is fraught with the danger of missing the true arguments of the global minimum of the function $|w(n)/\beta^{n-2}|$, since it is expected that this minimum is confined to a very small area. Table A2 shows an example in which the chosen grid $(\beta, w_0)$ contains the true arguments $(b, e(0))$.

Table A2. Estimates of $b$ and prediction of $Y(n+1)$; the true value are $b=2$; $e(0)=0.2$; we perfomed $m=1000$ simulations. The chosen grid of the pair $(\beta, w_0)$ contains the true arguments $(b, e(0))$.

|  | n=10 | n=15 | n=20 | n=25 |
|---|---|---|---|---|
| $\hat{\beta}$ | 2.20 ±0.78 | 2.03 ±0.29 | 2.00 ±0.0032 | 2.00 ±0.0001 |
| $\rho$ | 1.03 | 0.65 | 0.44 | 0.43 |
| REF | 19.1% | 3.9% | 0% | 0% |

One can observe that the quality of the estimation/prediction depends strongly on the sample size $n$. If $n$ drops down to $n \approx 10-12$, this quality quickly deteriorates. It should be noted that, for larger $b$'s, the quality of the estimation/prediction is getting better. E.g. for $b=3$; $n=15$; $e(0)=0.2$; we had:

(A16) $\qquad \hat{\beta} = 3.00 \pm 0.0001; \quad \rho = 0.32; \quad REF = 0\%.$

In the preceding examples, the grid over which the search was performed contained the true pair $(b, e(0))$. We now consider a grid which is *randomly shifted* with respect to the true pair $(b, e(0))$. For this purpose, we take the same fixed grid as above (with step $0.025$), but draw *e(0)* randomly in each simulation: *e(0) = ε N(0,1)*, where *N(0,1)* is a standard Gaussian random value, and ε is a small amplitude. The results of this experiment quantified by the pair *(ρ: relative improvement of the prediction, REF: fraction of rejected predictions)* are shown in Table A3 (*b=2*) and Table A4 (*b=3*).

Table A3. *Characteristics (ρ, REF) for random shifts of e(0) with respect to the grid; the true value for b is b=2. (ρ: relative improvement of the prediction, REF: fraction of rejected predictions). The parameter ε is defined through the expression e(0) = ε N(0,1), where N(0,1) is a standard Gaussian random value.*

| | n=15 | | |
|---|---|---|---|
| ε | 0 | $5 \cdot 10^{-7}$ | $10^{-6}$ |
| ρ | 0.53 | 0.54 | 0.64 |
| REF | 3% | 4% | 6.5% |
| | n=20 | | |
| ε | 0 | $2.5 \cdot 10^{-7}$ | $5 \cdot 10^{-7}$ |
| ρ | 0.45 | 0.51 | 0.65 |
| REF | 0% | 0% | 1% |
| | n=25 | | |
| ε | 0 | $5 \cdot 10^{-9}$ | $10^{-8}$ |
| ρ | 0.45 | 0.56 | 0.62 |
| REF | 0% | 0.5% | 1% |

Let us note that, as can be seen from equation (A1), the minimum possible value of the ratio *ρ* is:

(A17) $$\rho_{min} = (1 + b^2)^{-1/2}.$$

Thus, for *b=2*, this minimum value is *0.447*. Table A3 shows that, for *n=20* and *25*, this minimum value is reached for *ε = 0*, whereas for *n=15* it is not reached. What is particularly striking is that exceedingly small *ε*'s are sufficient to induce a significant deterioration of the prediction quality as measured by *ρ*. For *ε*'s larger than the values of the last column of Table A3, the ratio *ρ* becomes larger than 1, corresponding to a complete inefficiency of the prediction. It is interesting to note that, in such cases, the estimate of the parameter b still *can keep some efficiency,* but this does not translate into one for the prediction.

In table A4, we used a true value *b=3*, to illustrate that larger values of *b* require finer grids, as it was expected. We can therefore suggest the following *approximate, empirical* relation for the grid steps *Δb* and *Δe(0)*. When the grid steps *Δb* and *Δe(0)* obeys the following relation (A18), the global minimum of the function $|w(n)/ \beta^{n-2}|$ can in general be found with a good probability:

(A18) $$\Delta b = \Delta e(0) = 0.25/b^n.$$

This formula and the results presented in Tables A3 and A4 indicate that, for a practical implementation of the grid method, we are forced to take only restricted values of the sample size n (say, $n < 15$), and besides, the true value of the parameter $b$ should not be large (say, $b < 2.5$). Otherwise, the necessary very fine grid step would demand too much computation time.

Table A4. Characteristics *($\rho$, REF)* for random shifts of *e(0)* with respect to the grid; *b=3*.

| | n=10 | | |
|---|---|---|---|
| $\varepsilon$ | 0 | $5\cdot10^{-6}$ | $10^{-5}$ |
| $\rho$ | 0.64 | 0.66 | 0.85 |
| REF | 6.5% | 5% | 7.5% |

| | n=15 | | |
|---|---|---|---|
| $\varepsilon$ | 0 | $10^{-8}$ | $5\cdot10^{-8}$ |
| $\rho$ | 0.58 | 0.60 | 0.89 |
| REF | 4.5% | 6% | 13.5% |

| | n=20 | | |
|---|---|---|---|
| $\varepsilon$ | 0 | $5\cdot10^{-11}$ | $10^{-10}$ |
| $\rho$ | 0.51 | 0.58 | 0.61 |
| REF | 3.5% | 5% | 3.5% |

**Appendix B. Nonlinear auto-regression approach: The Nonlinear lag-operator**

The lag operator $L[e(t)] = e(t-1)$ is a very convenient tool for analyzing linear auto-regressive moving average (ARMA) models, see for instance (Box and Jenkins, 1976; Hamilton, 1994). In analogy with the linear lag operator, we introduce the non-linear lag operator B defined on functions f(t) with integer arguments:

(B1)    $B[f(t)] = f(t-1) f(t-2)$.

Then,

(B2)    $B^2[f(t)] = B[B[f(t)]] = B[f(t-1) f(t-2)] = f(t-2)f(t-3) f(t-3)f(t-4) = f(t-2)f^2(t-3) f(t-4)$.

For arbitrary positive integer powers of B, it can be shown that

(B3)    $B^n[f(t)] = \prod_{k=0}^{n} f^{C(n,k)}(t-n-k)$,

where $C(n,k)$ are the binomial coefficients. This allows us to use operator polynomials such as

(B4)    $1 + a_1 B + a_2 B^2 + \ldots + a_n B^n$,       $n > 0$,

where the $a_k$'s are arbitrary constants. Note the property

(B5)    $B^n[h f(t)] = h^{2n} B^n[f(t)]$,

which holds for an arbitrary constant $h$. Using the operator B, one can rewrite equation (1) in the following form:

(B6)    $r(t) = (1 + b B) e(t)$.

Let us apply the operator

(B7)    $1 - bB + b^3 B^2 - b^7 B^3 + \ldots + (-1)^k b^{g(k)} B^k + \ldots + (-1)^n b^{g(n)} B^n$;   $g(k) = k(k-1) + 1$;   $k = 1 \ldots n$.

to both sides of equation (B6). This yields

(B8)    $(1 - bB + \ldots + (-1)^n b^{g(n)} B^n) r(t) = (b + (-1)^n b^{g(n+1)} B^{n+1}) e(t)$.

Now, let us assume for a moment that $b^{g(n+1)} B^{n+1}[e(t)]$ tends to zero in some sense for $n \to +\infty$. Let us further assume that the coefficient $b$ is known. Then we would have from (B8) the limit construction rule for e(t)

(B9)    $e(t) = r(t) + \sum_{k=1}^{\infty} (-1)^k b^{g(k)} B^k [r(t)]$.

The relation (B8) would allow us to obtain $e(t)$ and $e(t-1)$ from all past values of $r(u)$, $u = t, t-1, t-2, \ldots$, and thus determine the best predictor

(B10)    $\hat{r}(n+1) = b\, e(n) e(n-1)$.

for *r(t+1)*. In the case where *b* is unknown, one could use its appropriate estimate, e.g. one obtained by the methods of statistical moments discussed in the main text for the case of Gaussian innovations *e(t)*. Even better, since the expectation of *e(t)* is zero, we can obtain a predictor for *r(t)* directly based on its past realizations without the need for inverting the innovations *e(t-1), e(t-1), ....* Indeed, let us replace *e(t)* by zero in the left-hand-side of (B9). This gives the equation

$$r(t) = - \sum_{k=1}^{\infty} (-1)^k b^{g(k)} B^k [r(t)]$$

which is nothing but a prediction of *r(t)* as a function of past realizations.

All this appears quite attractive, but these derivations are just formal manipulations and their justification needs a proof of convergence of series in equation (B8).

A fundamental problem in the nonlinear inversion that leads to (B9) is to determine under what conditions

(B11) $\quad b^{g(n)} B^n [e(t)] = b^{g(n)} \prod_{k=0}^{n} [e(t-n-k)]^{C(n,k)} \to 0 \quad$ as $n \to +\infty$.

The product contains *n+1* terms *e(u)*'s with powers *C(n,k)* running from 1 to *C(n,Int[n/2])* $\approx (2/\pi n)^{1/2} 2^n$ (by the Stirling formula), where *Int[n/2]* is the integer part of *n/2*. An obvious sufficient condition for the validity of (B11) is that |*e(u)*| be bound from above by *exp(−ε)*, with *ε<0*. Then, the term (B12) is bound from above by *exp(-ε $2^n$)* and the term (B11) converges to 0 for arbitrary values of *b* (even for |*b*|>*1*). When |*e(u)*| is not bounded from above by *1*, the situation is more subtle. Let us assume that the *e(u)*'s are standard i.i.d. Gaussian random values. We are going to show that in such a case the expectation of the absolute value of (B11) tends to infinity for any |*b*| > *0*. The moments of *e = e(u)* are the following:

(B12) $\quad E[e^{2r}] = (2r)!/(2^r r!); \quad E[|e|^{2r+1}] = \pi^{-1/2} 2^{r+1/2} Gamma(r+1); \quad r = 0,1,2,...$

Using these formulae, it is easy to show that

(B13) $\quad E |b^{g(n)} B^n [e(t)]| = b^{g(n)} \prod_{k=0}^{n} E[|e^k|] > b^{n(n-1)+1} exp(\delta 2^n),$

where *δ* is some positive constant. Thus, (B13) tends to infinity for any positive |*b*|.

It is perhaps possible to develop some regularization procedure of the process *r(t)* to taper its large values and thus providing convergence of the series in (B9) but this question requires more careful investigation. Thus, the only known class of distributions so far that can guarantee convergence of (B9) is the class of distributions bounded from above. This class seems rather restrictive though we can note that, as it was recently shown (*Pisarenko et al., 2007*), such distributions bounded from above can model populations with seemingly "heavy tail" behavior, such as earthquakes. The third Gumbel or the Weibull distribution provides an example in the real of extreme value distributions. Thus, the technique suggested in this Appendix can be used with such distributions.